Costantino Sigismondi
*Storia dell'Astronomia, Università di Roma "La Sapienza"*
sigismondi@icra.it

# Moti e distanze angolari in cielo con telescopio e cronometro

**Abstract:** A basic telescope and a chronometer can be used to learn classical methods of spherical astronomy. The conjunction Venus-Saturn of July 2, 2007 has been observed through July 3 with a 50 mm telescope.
Planetary daily motions and angular distances with respect to ψ Leonis, have been obtained with telescope field of view crossing times with an accuracy of 1 arcminute. In the age of CCD astronomy and planetarium programs for personal computer available to everyone, those are educational occasions to approach 3-D astronomy from real data, self obtained.

**Introduzione**

La recente congiunzione Saturno-Venere del 2 luglio 2007 ha riproposto l'occasione di confrontarsi con l'astronomia sferica, ossia la misura di distanze e moti sulla sfera celeste. L'esame quantitativo dei moti orbitali di Venere e di Saturno, visti da Terra, rappresenta un'esperienza scientifica di grande valore, poiché ci aiuta a comprendere meglio il lavoro degli astronomi del passato e del presente e a toccare con mano la complessità dell'analisi dei dati, non automatizzata mediante software preconfezionati.

Il presente articolo vuole invitare i docenti a sfruttare queste occasioni per imparare insieme agli studenti che con un cronometro ed un piccolo telescopio o binocolo, si può conseguire la precisione di un arcominuto nelle misure di posizione relativa degli astri, quella raggiunta da Tycho Brahe e i suoi astronomi con l'occhio nudo ad Uraniborg, il primo osservatorio moderno.

Inquadrando un oggetto nel telescopio, senza usare il motore di inseguimento, questo si inizia a muovere verso Ovest con un angolo rispetto all'orizzontale che va approssimativamente da +45° a -45° passando per 0° quando l'oggetto passa al meridiano. Queste vengono chiamate anche misure di drift.

L'uso del Bright Stars Catalogue, il catalogo stellare delle stelle visibili ad occhio nudo, e di Ephemvga, un programma di effemeridi, completano l'osservazione in fase di progettazione ed analisi dati.

Venere ha avvicinato ψ Leonis e Saturno il 2 luglio, e le misure rispetto a questa stella fissa aiutano a capire i moti dei due pianeti separatamente. L'identificazione di ψ Leonis è stata possibile proprio con l'uso incrociato del catalogo, delle effemeridi planetarie e dei tempi di volo dei pianeti e della stella nel campo di vista del telescopio.
Il lavoro si articola nel seguente modo:

1) Misura del campo di vista del telescopio
2) Differenza in ascensione retta tra Venere e Saturno
3) Calcolo della loro differenza in declinazione
4) Misure rispetto a ψ Leonis

**Campo di vista del telescopio**

Il campo circolare che appare al telescopio ha una dimensione angolare che deve essere misurata per procedere alle misure che ci interessano.

Per farlo conviene selezionare un oggetto di cui si conosca già la declinazione e lo si pone al centro del campo di vista. Per identificare il centro del campo di vista si può fare ad occhio, oppure, meglio, incrociando due capelli attaccati ai bordi dell'oculare (con oculari Kellner i capelli vengono messi nella parte che va dentro il telescopio), o ancora scegliendo stelle che transitano al meridiano e che quindi si muovono, in quel momento, in orizzontale.

Una volta nota la declinazione δ dell'astro posto al centro del campo di vista (CdV) il tempo τ di uscita dell'astro dal CdV si trasforma nel raggio del CdV in minuti d'arco con la formula:

$$rCdV['] = \tau[s] \cdot \cos(\delta)/4$$

Questa formula deriva dalla proporzione:
360° : 24h = x° : t
Dove t è il tempo impiegato a percorrere il CdV. Dalla proporzione si ricava infatti che:
x' = (360*60)*t(s)/(24*3600)
da cui si ottiene appunto x'=t/4=0.25t.

La formula non è quindi mai precisa in quanto si considera che un giro di 360° su un parallelo celeste viene fatto in 24 ore cosa che non accade mai (per le stelle è sempre di 23h e 56m 04s, per la Luna è variabile e tra il 30 giugno ed il 1 luglio 2007 ha impiegato 24h 56 m).

t/4 è poi moltiplicato per il coseno della declinazione, che tiene conto dell'assottigliamento dello spicchio di meridiano muovendosi verso i poli.

Venere in quei giorni tornava in meridiano ogni 23h 59 m, quindi l'uso della formula approssimata genera degli errori, minimi per stelle e pianeti, ma fino al 4% per la Luna.

Con un telescopio TASCO zoom focus 15x 45x, fissato a 45x ho provato il metodo della stella al centro e quello della stella lungo un diametro orizzontale (il più facile da identificare ad occhio). E' risultata utile pure la Luna piena, posta al centro del campo, più facile da centrare rispetto ad una stella. Ecco i risultati per valutare gli errori di misura.

**Tab. 1 Misura del Campo di Vista (formule esatte)**

| Oggetto | τ[s] | δ[°] | ØCdV['] |
|---|---|---|---|
| Venere [al centro] | 184 x 2 | +14°53' | Ø=88'.98 |
| Luna piena [al centro] | 202 x 2 | -27°07' | Ø=86'.98 |
| Antares [in meridiano] | 381 | -26°27' | Ø=85'.52 |
| Giove [in meridiano] | 372 | -21°35' | Ø=86'.74 |

Si noti che mentre per pianeti e stelle si è misurato il tempo di volo dentro al campo di vista, per la Luna Piena si è dovuta prendere la media del tempo di primo contatto col bordo del CdV e di ultimo contatto per valutare l'istante in cui il centro della Luna usciva fuori dal CdV.
Se la Luna non fosse stata piena questo metodo non si sarebbe potuto applicare.
Nel caso degli oggetti posti al centro è possibile che non si trovassero esattamente al centro e la misura del diametro del CdV sia stata così stimata per eccesso o per difetto; mentre nel caso dei passaggi al meridiano, con l'astro che si muove orizzontalmente, trattandosi di misure del diametro, è sempre possibile individuare una corda leggermente più corta del diametro. Come valore rappresentativo prendo media e varianza dei quattro dati 87'±1'.4. Scartando il dato di Venere -che evidentemente non era esattamente al centro del CdV- si ha 86'.4±0.8'.

### Differenza di ascensione retta tra Venere e Saturno

L'oggetto descrive una corda nel CdV, ed il tempo intermedio tra ingresso ed uscita nel CdV è quello in cui l'oggetto è in asse con il centro del CdV. Un altro oggetto entra ed esce dal CdV in tempi differenti, il confronto tra i tempi in cui i due oggetti sono in asse con il centro del CdV dà la loro differenza in ascensione retta.

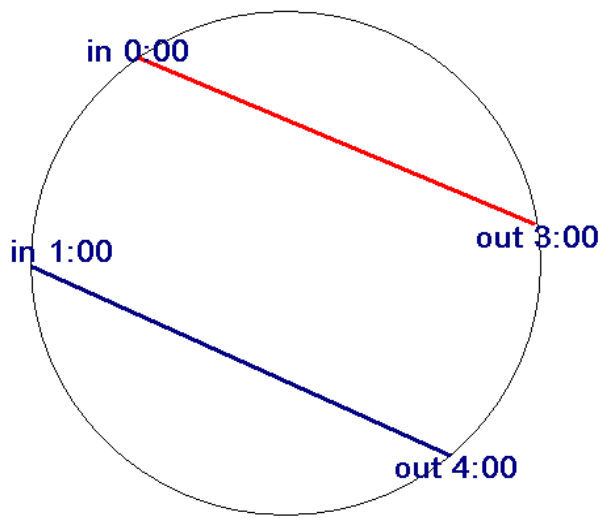

Nell'esempio in figura 1 il cronometro parte quando l'oggetto rosso entra nel CdV. Il rosso è in asse col centro dopo 1:30, mentre il blu dopo 2:30, la loro differenza in ascensione retta è di 1 minuto, maggiore per il blu.

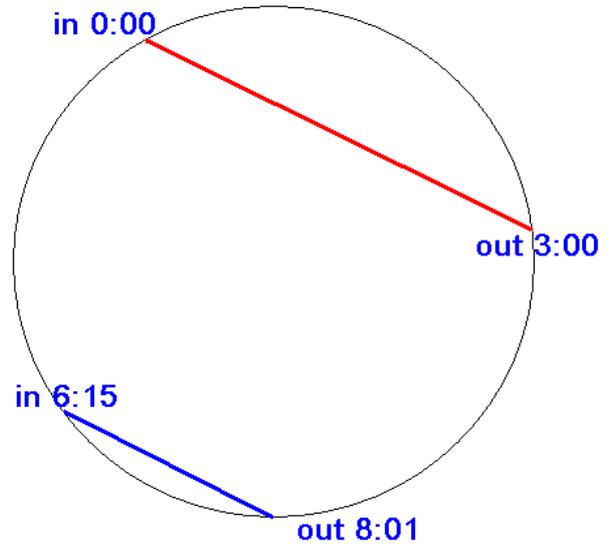

In figura 2 è il caso in cui l'oggetto blu entri nel CdV ben dopo che quello rosso ne sia uscito. Il cronometro continua ad andare e la media per i tempi del blu dà 7:08, 5 minuti e 38 secondi dopo la media del rosso. Questa è la differenza di ascensione retta tra i due oggetti. L'abilità sta nel puntare il rosso in modo tale che il blu entri nello stesso campo di vista dopo il tempo necessario.

Ecco le tabelle delle osservazioni dal 30 giugno al 3 luglio, comprensive della sincronizzazione col tempo universale e di quando gli oggetti sono in asse con il CdV.

**Tab. 2 Osservazioni**

| Tempo del cronometro | 30 giugno 2007 - Evento |
|---|---|
| 0:00 | Venere entra nel CdV |
| 1:26 | Saturno entra |
| 3:10 | Venere esce |
| 5:14 | 19:44:00 UTC |
| 7:06 | Saturno esce |
| 1:35 | Venere in asse |
| 4:16 | Saturno in asse |

| Tempo del cronometro | 1 luglio 2007 - Evento |
|---|---|
| 0:00 | Venere entra nel CdV |
| 1:08 | Saturno entra |
| 5:34 | Saturno esce |
| 5:48 | Venere esce |
| 6:21 | 19:48:00 UTC |
| 2:54 | Venere in asse |
| 3:21 | Saturno in asse |
| Tempo del cronometro | 2 luglio 2007 - Evento |
| 0:00 | Saturno entra nel CdV |
| 0:12 | Venere entra |
| 2:11 | Saturno esce |
| 2:18 | ψ Leonis entra |
| 5:25 | Venere esce |
| 6:00 | ψ Leonis esce |

| 6:25 | 20:08:00 UTC |
|---|---|
| 2:48.5 | Venere in asse |
| 1:05.5 | Saturno in asse |
| 4:09 | ψ Leonis in asse |

| Tempo del cronometro | 3 luglio 2007 - Evento |
|---|---|
| 0:00 | Saturno entra nel CdV |
| 1:10 | ψ Leonis entra |
| 1:22 | Saturno esce |
| 2:59 | Venere entra |
| 5:27 | ψ Leonis esce |
| 6:01 | Venere esce |
| 6:43 | 20:28:00 UTC |
| 4:30 | Venere in asse |
| 0:41 | Saturno in asse |
| 3:18.5 | ψ Leonis in asse |

Sia il 30 giugno che il primo luglio Saturno ha un'ascensione retta maggiore di quella di Venere, poi la situazione si inverte. Possiamo calcolare l'istante della congiunzione, quando le ascensioni rette sono uguali.

**Tab. 3 Differenze in ascensione retta**

| Data UT | Saturno- Venere | [s] |
|---|---|---|
| Giu 30.822 | 2 m 41 s | 161 |
| Lug 1.825 | 0 m 27 s | 27 |
| Lug 2.839 | -1 m 43 s | -103 |
| Lug 3.853 | -3 m 49 s | -229 |

Interpolando dalla tabella 3 l'istante per cui le ascensioni rette si uguagliano, la congiunzione risulta avvenuta il 2 luglio alle 0:51 Tempo Universale, in ottimo accordo con le ore 0h 42 del TU che vengono dalle effemeridi.

**Differenza di declinazione tra Venere e Saturno**

Confrontando i tempi di attraversamento delle corde del CdV con quello che, alla stessa declinazione, occorrerebbe per percorrere il diametro di 86.4', si calcola immediatamente la distanza delle corde dal centro del CdV.
Assumiamo, per tutti i calcoli seguenti, la declinazione fissa di +14°, che è quella di ψ Leonis.
Il diametro del CdV viene percorso in
τCdV=86.4 ·4/cos(14°)=356.4 s.

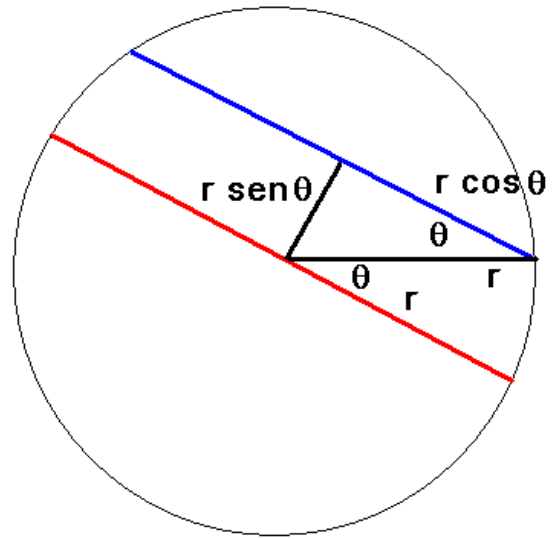

Ponendo r=43.2' nella figura 3, la distanza della corda percorsa in τc dal centro del CdV è pari a
dc=43.2'·sin[arccos(τc/356.4)]
Queste distanze vengono calcolate sia per Venere che per Saturno che per ψ Leonis e poi sommate algebricamente tra loro per ottenere le distanze relative. Il segno negativo indica che l'oggetto è transitato nella metà inferiore del CdV, il positivo in quella superiore.

**Tab. 4 Tempi di attraversamento del Campo di Vista**

| Data | Venere τc | Saturno τc | ψ Leonis τc |
|---|---|---|---|
| Giu 30.822 | -190 | -340 | --- |
| Lug 1.825 | -348 | 226 | --- |
| Lug 2.839 | -313 | 131 | -222 |
| Lug 3.853 | -182 | 82 | -257 |

**Tab. 5 Distanze corda - centro del Campo di Vista**

| Data\ dc['] | Venere | Saturno | ψ Leonis |
|---|---|---|---|
| Giu 30.822 | -36'.5 | -12'.9 | --- |
| Lug 1.825 | -9'.3 | 33'.4 | --- |
| Lug 2.839 | -20'.7 | 40'.2 | -33'.8 |
| Lug 3.853 | -37'.1 | 42'.0 | -30'.0 |

Per il 2 e 3 luglio abbiamo il riferimento della stella fissa ψ Leonis.
Scegliendo questa stella come origine del sistema di riferimento abbiamo per questi due giorni le seguenti posizioni:

**Tab. 6 Distanze relative a ψ Leonis**

| Data\ dec['] | Venere | Saturno | ψ Leonis |
|---|---|---|---|
| Lug 2.839 | 13'.1 | 74'.0 | 0' |
| Lug 3.853 | -7'.1 | 72'.0 | 0' |
| Data\ r.a.[s] | Venere | Saturno | ψ Leonis |
| Lug 2.839 | -80.5 | -183.5 | 0 |
| Lug 3.853 | 71.5 | -157.5 | 0 |
| Data\ r.a.['] | Venere | Saturno | ψ Leonis |
| Lug 2.839 | -19'.5 | -44'.5 | 0 |
| Lug 3.853 | 17'.4 | -38'.2 | 0 |

Per l'ultima serie di dati le unità di misura sui due assi sono uguali, avendo moltiplicato per cos(δ) con δ=14°.

Il moto angolare totale di Venere risulta così di 42.1'/giorno mentre quello di Saturno di 6.61'/giorno, e sono rappresentati in figura 4.

**Identificazione della stella di riferimento con effemeridi e catalogo**

L'identificazione di una stellina di magnitudine 5.35 comparsa nella luce crepuscolare avendo come punti di riferimento solo i due pianeti si fa con le coordinate equatoriali relative.

Dalle effemeridi [Ephemvga] sappiamo che al momento delle due osservazioni Venere e Saturno avevano coordinate [relative all'equinozio 2007.5] come in tabella:

**Tab. 7 Effemeridi per Venere e Saturno**

| Data | Ven α | Ven δ | Sat α | Sat δ |
|---|---|---|---|---|
| 2 Lug | 9h42.9m | 14°11' | 9h41.2m | 15°12' |
| 3 Lug | 9h45.4m | 13°50' | 9h41.6m | 15°10' |

Ephemvga, un software molto utile per calcoli astronomici, consente di impostare l'epoca con una precisione del decimo di anno, nel caso in questione 2007.5 corrisponde casualmente proprio al 2.875 luglio. In altri casi l'errore che si commette calcolando la precessione degli equinozi è comunque inferiore a 0.1' [0.06 m di ascensione retta].

Sottraendo le distanze relative con la stella di campo ricavate in tab. 6 si ottengono le sue coordinate equatoriali (epoca 2007.5).

**Tab. 8 Coordinate di ψ Leo derivate dalle osservazioni**

| Data | daVen α | daVen δ | daSat α | daSat δ |
|---|---|---|---|---|
| 2 Lug | 9h44.24m | 13°57'.9 | 9h44.26m | 13°58'.0 |
| 3 Lug | 9h44.21m | 13°57'.1 | 9h44.22m | 13°58'.0 |

Nel Bright Stars Catalogue in corrispondenza dell'ascensione retta 9h 44m 15s per l'equinozio 2000.0 c'è solo la stella 3866 ψ Leonis di m=5.35 compatibile con quel valore della declinazione.

Le coordinate di ψ Leonis all'equinozio 2007.5 sono:
AR 9h 44m 08.4s e declinazione 13° 59' 13" [Astronomical Almanac] Ephemvga arrotonda a α=9h44.1m e δ=13°59'.

I valori medi della tabella 8 danno α=9h44m14s e δ=13°57'45", con scarti massimi di circa 1.5 minuti d'arco. Il confronto con i dati in tabella ci dà anche la precisione del metodo adoperato: 1'.5 in declinazione e 5.6 s in ascensione retta (1'.4).

L'errore sulle coordinate equatoriali assolute dipende principalmente dalla difficoltà di osservare la stella di m=5.35 nella luce crepuscolare e a 6° sull'orizzonte ai bordi del campo di vista. L'incertezza su alcune misure è stata dell'ordine dei 5 secondi, laddove per i pianeti non superava mai 1 secondo.

**Altro metodo per misurare il diametro del Campo di Vista**

Osservando due stelle con piccola differenza in declinazione attraversare il CdV, dalla misura dei tempi di attraversamento si può ricavare il raggio del CdV invertendo la formula

Δδ=dc1± dc2=
=r'·sin[arccos(τc1/τd)] ± r'·sin[arccos(τc2/τd)]
sapendo che r'= τd·[s] cos δ/4

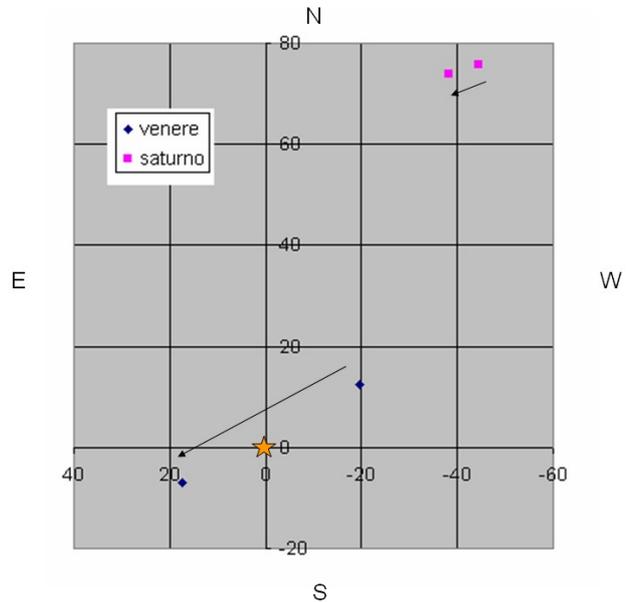

La figura 4 mostra il moto dei due pianeti rispetto a ψ Leonis dal 2 al 3 Luglio 2007.

*Bibliografia*